\documentclass[review]{elsarticle}
\usepackage[utf8]{inputenc}

\usepackage{lineno,hyperref}
\usepackage{amsmath}
\usepackage{amsfonts} 
\usepackage{longtable}
\usepackage[official]{eurosym}
\usepackage[dvipsnames]{xcolor}
\modulolinenumbers[5]

\journal{Applied Energy}

\begin{document}

\begin{frontmatter}

\title{Assessing the impact of inertia and reactive power constraints in generation expansion planning}


\author[mymainaddress]{S. Wogrin\corref{mycorrespondingauthor}}
\cortext[mycorrespondingauthor]{Corresponding author:}
\ead[url]{sonja.wogrin@comillas.edu}

\author[mysecondaryaddress,mymainaddress2]{D. Tejada-Arango}
\author[mythirdaddress]{S. Delikaraoglou}
\author[mythirdaddress]{A. Botterud}

\address[mymainaddress]{Institute for Research in Technology (IIT), School of Engineering (ICAI), Comillas Pontifical University, Madrid, Spain}
\address[mysecondaryaddress]{Global Trading - Fundamentales y Competencia, Endesa SA, Madrid, Spain}
\address[mymainaddress2]{Faculty of Economics and Business Administration (ICADE), Comillas Pontifical University, Madrid, Spain}
\address[mythirdaddress]{Laboratory for Information and Decision Systems (LIDS), Massachusetts Institute of Technology, Cambridge, MA}

\begin{abstract}
On the path towards power systems with high renewable penetrations and ultimately carbon-neutral, more and more synchronous generation is being displaced by variable renewable generation that does not currently provide system inertia nor reactive power support. 
This could create serious issues of power system stability in the near future, and countries with high renewable penetrations such as Ireland are already facing these challenges. Therefore, this paper aims at answering the questions of whether and how explicitly including inertia and reactive power constraints in generation expansion planning would affect the optimal capacity mix of the power system of the future. Towards this end, we propose the novel Low-carbon Expansion Generation Optimization (LEGO) model, which explicitly accounts for: unit commitment constraints, Rate of Change of Frequency (RoCoF) inertia requirements and virtual inertia provision, and, a second-order cone programming (SOCP) approximation of the AC power flow, accounting for reactive power constraints. An illustrative case study underlines that disregarding inertia and reactive power constraints in generation expansion planning can result in additional system cost, system infeasibilities, a distortion of optimal resource allocation and inability to reach established policy goals. 
\end{abstract}


\end{frontmatter}


\section{Introduction}
\label{sec:Intro}

\subsection{Motivation}
\label{subsec:Motivation}

The IPCC Global Warming report \cite{IPCC} urges us to halve CO$_2$ emissions by 2030 in order to avoid devastating changes to our climate. As a consequence, there have emerged worldwide ambitious policy targets such as: the American Clean Energy and Security Act \cite{congress2009american}; the 2030 climate and energy framework \cite{EU2030} that was adopted by the European Union in 2014 with goals to cut 40\% of greenhouse gas emissions, to achieve a 32\% share of renewable energy and a 32\% improvement in energy efficiency by 2030; or, the European Commission’s goal to be carbon-neutral by 2050 \cite{EC2050}. One crucial step in all these plans is the decarbonization of the electric power sector. However, it is important to understand that replacing traditional thermal generation provided by synchronous machines with largely intermittent renewable energy sources requires a fundamental paradigm change of how our power systems are being operated. This underlying paradigm change entails many technical difficulties and challenges, such as reactive power and inertia support \cite{rezkalla2018electric}, that will have to be overcome before the goal of a completely carbon-neutral power system can be achieved. 

While reactive power and system inertia do not seem to be an issue in most power systems currently, their relevance for the smooth and reliable functioning of the system is indispensable \cite{tielens2016relevance,NREL2020}. Moreover, in a carbon-free power system where most thermal dispatchable plants have been shut down, reactive power and inertia support become an important issue when it comes to system stability and security. As a matter of fact, inertia is already becoming an issue in countries with high renewable penetration, such as Ireland. EirGrid, the transmission system operator of Ireland has proposed an explicit rate of change of frequency (RoCoF) requirement in the grid code to facilitate the delivery of the 2020 renewable targets \cite{CER2014}.

While the impact of reactive power and inertia have been studied extensively in operational problems \cite{Daly15,Paturet2020}, they have been largely ignored in generation expansion planning (GEP). This paper aims to develop a Low-carbon Electricity Generation Optimization (LEGO) model, which allows us to assess how the optimal generation expansion plan for the power system of the future would change if both system inertia via RoCoF and reactive power constraints are accounted for explicitly. We consider reactive power constraints in an AC optimal power flow (OPF) setting approximated by second-order cone programming (SOCP). We furthermore account for realistic unit-commitment type operating constraints such as start-up, shut-down, or ramping constraints, and a novel formulation that integrates RoCoF within GEP.

\subsection{Literature Review}
\label{subsec:LitRev}

This literature review is by no means an exhaustive overview over all different generation expansion planning (GEP) approaches. As a matter of fact, there exist many interesting GEP review papers that discuss a plethora of important aspects for GEP of the future power system. Koltsaklis and Dagoumas \cite{koltsaklis2018state} categorize GEP approaches with respect to economic, environmental, regulatory, and technical aspects. Babatunde et al \cite{babatunde2019comprehensive} classify GEP models with respect to how they treat the time horizon, uncertainty, what market structure they assume, and their network topology, and mention that only 19\% of the analyzed publications explicitly consider a network. Among other topics, they also focus on representation of unit commitment (UC) details and storage representation within GEP models. For further details about generation and transmission planning in deregulated power markets, the reader is referred to Gonzalez et al \cite{gonzalez2020review}. 

The focus of the literature review carried out in this paper lies on realistic GEP and UC models that include the network topology, an explicit representation of renewable energy sources, battery energy storage systems (BESS) with degradation, and that additionally account for: RoCoF inertia constraints and reactive power provision. Palmintier and Webster \cite{Palmintier2011} have already pointed out the importance of considering UC constraints in generation expansion planning with high penetration of renewables. However, they do not consider the network, reactive power or inertia. Since there is no single work that covers all of these topics at once - this is the research gap that we are trying to fill with this paper - we discuss the most relevant works by topic.

Let us first focus on the topic of inertia.
Inertia modeling in unit commitment setting has already been discussed in several previous studies. The authors in \cite{EnergiforskReport,perez2016robust,daly2015inertia} incorporate the RoCoF constraint in the unit commitment problem, using the swing equation of  Center-of-Inertia  (CoI). Aiming to include frequency deviation metrics, the work in \cite{Ahmadi2014} includes an analytic formulation of frequency nadir and limits the maximum post-disturbance frequency deviation from the nominal set point, caused by an instantaneous load increase. Following an alternative approach that does not require the explicit modeling of turbines and decouples governor control from system frequency, papers \cite{Badesa2018,Brito2018} employ a simplified analytical formulation that requires strict assumptions on system damping and nodal frequency response provision. Using an analogous approach, \cite{Trovato2019} co-optimize the energy generation schedule and the provision of frequency response reserves. The interaction of stochastic renewable energy resources with inertia constraints and their combined impact on generation scheduling is studied in \cite{Teng2016StochasticSW} using a stochastic unit commitment problem formulation. The work in \cite{paturet2020stochastic} extends this modeling framework, including equipment contingencies and using analytic expressions of relevant  frequency  metrics  as  functions  of  the  system variables (e.g., inertia, damping, aggregate droop gain) that are computed endogenously in the stochastic unit commitment problem. 
An important outcome of \cite{paturet2020stochastic} is that the limit on maximum instantaneous RoCoF is typically the most restrictive constraint in an inertia-aware unit commitment problem, compared to the limits on frequency nadir and on quasi steady-state frequency deviation. Therefore, in this paper we focus on the impact of RoCoF constraint requirements on the optimal expansion planning schedule.

Regarding literature that covers inertia modeling in a generation expansion planning framework, the most relevant studies focus on the optimal sizing and placing of virtual inertia in the power system. The authors in \cite{poolla2017optimal} formulate a robust inertia allocation problem that finds the optimal placing of virtual inertia accounting for the worst-case disturbance location, whereas \cite{poolla2019placement} considers the optimization of geographical dispersion and parameter tuning of grid-following and grid-forming virtual inertia devices used for inertia emulation to improve the resilience of low-inertia grids. In addition, \cite{markovic2019optimal} presents a method for optimal sizing of storage capacity in terms of power to energy capacity ratio and tuning of virtual inertia and damping gains of the associated grid-forming converters in order to ensure sufficient energy and power capacity for meeting a predefined active power imbalance. All the aforementioned works focus primarily on the technical aspects of new virtual inertia installations, without considering the economic impact of those investments and their interplay with the other power production technologies. To this end, this work aims at formulating an inertia-aware generation expansion planning model that captures the complementarity between inertia and energy services in a unified framework that accounts for the technical but also economic impacts of inertia provision in systems with high renewable penetration.

Next, let us focus on reactive power constraints in an AC-OPF framework with UC constraints. The AC-OPF allows us to represent both active and reactive power constraints in optimization models for electricity systems. Due to the non-convex nature of these constraints, the optimization process becomes a Non-Linear Programming (NLP) problem that is hard to solve. On top of that, UC models introduce binary variables to the optimization problem, making the UC AC-OPF a non-convex Mixed-Integer Non-Linear Programming (MINLP) problem - one of the most difficult types of problems in the literature \cite{Tang2015}. It is the combination of UC and AC-OPF constraints that makes it so challenging to solve, and hence, finding the global optimum cannot be guaranteed \cite{Castillo2016}. In order to overcome this issue, it is common to apply approximations to the non-convex AC-OPF constraints, e.g., DC-OPF, Second-Order Cone Programming (SOCP), or other solution techniques \cite{TejadaArango2019}.

On the one hand, the DC-OPF constraints linearize the active power flow equation, while disregarding the reactive power component (i.e., assuming that all voltages are equal to 1 p.u.). This is a well-known approach for GEP and TEP problems that also involve UC constraints, e.g., GEP-UC \cite{TejadaArango2020}, TEP-UC \cite{Golestani2010}, and GEP-TEP-UC \cite{Guerra2016}. However, the main drawback appears when voltages are relevant in the optimization, e.g., a high-cost generation unit must run to support the voltage in a specific bus with its reactive power. Such a case can simply not be captured under a DC-OPF framework.

On the other hand, the SOCP is one of the convexification methods that approximates the AC-OPF constraints \cite{ZOHRIZADEH2020}, including the reactive power equations \cite{Low2014a} and \cite{Low2014b}. The SOCP approximation with UC constraints leads to a Mixed-integer quadratic programming (MIQP) problem, which has the advantage that commercial solvers can solve it and find the globally optimal solution. However, as in the DC-OPF, the SOCP approximation can lead to solutions that are not AC-feasible. By introducing additional constraints in the SOCP formulation, the approximation error can be reduced \cite{Kocuk2016} which helps to retrieve an AC-feasible solution. Despite these advantages, the SOCP has barely been used in TEP problems \cite{GHADDAR2019}, and we are not aware of GEP/TEP problems with UC constraints using the SOCP approximation. In the literature, we have found research on GEP/TEP using AC-OPF constraints, as \cite{KOLTSAKLIS2018} shows in its review of the state-of-the-art, however, they do not include UC constraints. Moreover, as they consider the full AC-OPF (and no convexification such as the SOCP), they have the non-convex difficulties of MINLP problems.
To the best of our knowledge, there is a research gap for a GEP model that considers both the UC constraints and the AC-OPF, which is what we address with the proposed LEGO model.

Therefore, the original contributions of this work are 1) the formulation of the LEGO model itself as being - to the best of our knowledge - the first generation expansion model in the literature that simultaneously considers UC, AC-OPF and reactive power constraints (approximated by SOCP), and RoCoF inertia requirements considering both virtual inertia and inertia provided by synchronous machines; and, an in depth analysis of the impact of inertia and reactive power constraints on GEP for different renewable penetrations.

The remainder of the paper is organized as follows: section \ref{sec:Methodology} contains the mathematical formulation of the LEGO model; in section \ref{sec:Results} we present numerical results that showcase the impact of inertia and reactive power on GEP decisions. Finally, section \ref{sec:Conclusions} concludes the paper.

\section{Low-carbon Electricity Generation Optimization Model}
\label{sec:Methodology}

\allowdisplaybreaks

This section contains the mathematical formulation of the novel Low-carbon Expansion Generation Optimization (LEGO) model, which - as its acronym implies - is designed in a modular fashion to maximize model flexibility. The LEGO model is flexible in two aspects: in terms of how time is represented; and, in terms of thematic modeling blocks that can be combined among each other. 

In section \ref{subsec:ModelTime} we explain the representation of the time horizon in the LEGO model. Then, we present the model formulation, and each of the individual LEGO blocks, being: standard constraints (objective function, UC, storage, renewable and DC-OPF constraints) in section \ref{subsec:ModelStandard}. Section \ref{subsec:ModelInertia} focuses on deriving the novel way to consider inertia, and in particular RoCoF, constraints in an expansion planning framework. In section \ref{subsec:ModelReactivePower} we present the SOCP approximation of the AC-OPF LEGO block. Finally, section \ref{subsec:OverviewModels} provides an overview of how the different LEGO blocks can be assembled to carry out different case studies.

\subsection{Representation of Time}
\label{subsec:ModelTime}

There are several different approaches for representing time in generation expansion planning models: the exact representation of each individual hour of the time horizon to be studied, which is usually computationally intractable; some kind of representative periods such as days or weeks that adequately represent the time horizon \cite{reichenberg2018policy}; or, representing the time horizon through multi-hour time slices (often referred to as time blocks, time periods) that could be chronological or not depending on the approach \cite{Wogrin11}. Each of those approaches have their pros and cons. In this paper, however, we want to present a flexible model formulation that allows us to pick either of those methods, and not having to choose only one of them.

To that purpose, we introduce three different temporal indices: $p,k, rp$, which we will use throughout the model formulation. Index $p$ represents the actual chronological periods (which are usually hourly); $rp$ are the representative periods used; and finally, $k$ correspond to the chronological periods within the representative period $rp$. We also introduce the parameter $W^{RP}_{rp}$, which represents the weight of this representative period. Parameter $W_k^K$ is the weight of the period $k$ within each $rp$. Finally, there is also a mapping $\Gamma(p,rp,k)$ that relates each actual period $p$ to its representative period $rp$ and period $k$.

Let us demonstrate the flexibility of this notation by presenting a simple example. Imagine we have one year's worth of hourly data available, and we want to run our model in two different ways: a) the exact chronological hourly model; and b) representative days approximation of the original data with 7 representative days. By a simple adjustment of the temporal indices and the weights we can run both options a) and b) without having to change the model formulation. For both options we would have 8760 chronological periods $p$. 

For the exact hourly model (option a), we simple set index $rp$ to one. We only have one representative period, which is the year itself. Index $k$ are the chronological periods within the year, so $k$ ranges from 1 to 8760 and has the same cardinality as $p$ in this case. All the weights, both $W^{RP}_{rp}$ and $W_k^K$, are equal to 1 in option a). And $\Gamma$ in this case simply associates $p$ with the corresponding $k$.

In option b) we represent one year of data by using 7 representative days $rp$. Then $W^{RP}_{rp}$ simply states how many actual days (out of the 365) are being represented by one of the 7 representative days $rp$ \footnote{The sum of $W^{RP}_{rp}$ over all representative day is always 365.}. These numbers could be the result of a clustering algorithm \cite{omran2007overview}. Since the representative period chosen is one day, $k$ ranges from 1 to 24, i.e. the 24 hours that represent each representative day. Since we are still sticking to hours, each of the 24 weights $W_k^K$ is still 1. 
The mapping $\Gamma$ is a little bit more complicated, but could also stem from a clustering algorithm. Let us consider January 1st: imagine that this day is represented by $rp_5$ for example, then $\Gamma(p_1,rp_5,k_1)=1 \dots \Gamma(p_{24},rp_5,k_{24})=1$. January 2nd is represented by $rp_2$ for example, so the 25th hour of the year would lead to a $\Gamma(p_{25}, rp_2, k_1)=1$, and so on.

With this in mind, we quickly want to define the notation of double minus $--$ or double plus $++$ that appears sometimes in the remainder of this section. The term $k--1$ simply refers to the previous within-time period $k$. For example, if $k=2$, then $k--1$ corresponds to $k=1$. But, if $k=1$, then $k--1$ corresponds to $k=24$. The double minus creates a cyclic link between the first and the last $k$ of the same representative period. In the remainder of the paper, we use this terminology for commitment variables and for cyclic storage constraints.

Finally, while this temporal structure might seem convoluted at first sight, it is a powerful tool that allows us to maintain the highest degree of model flexibility and versatility. In any case and for the sake of simplicity, one can always think of an hourly chronological model, as described by option a), and move on.

\subsection{Standard Constraints}
\label{subsec:ModelStandard}

Section \ref{subsec:ModelStandard} contains the constraints of the generation expansion model that can be considered standard in this type of literature. Since they do not represent an original contribution, nor a novelty, they will be discussed only briefly.

The full notation of all model indices, parameters and variables can be found in the appendix. But for the sake of clarity, index $g$ represents all generating units as a whole (both existing and candidate units), and sub-indices $t,r,s$ are thermal, renewable and storage units.

The objective function \eqref{eqn:ObjectiveFunction} represents total system cost as: thermal production cost (start-up cost, commitment cost, and variable cost); potential cost for non-supplied energy;
cost of providing upward and downward secondary reserves by thermal and storage units; finally, investment costs for building new units.
Constraint \eqref{eqn:PNS} represents the upper and lower bounds of non-supplied energy; and, \eqref{eqn:BoundInvestments} defines investment variables as non-negative integers and establishes an upper bound introduced by parameter $\overline{X}_g$.

\begin{subequations}
\label{eqn:GeneralConstraints}
\begin{align}
min  \sum_{rp,k}  W^{RP}_{rp}  W^{K}_{k} \Big(   \sum_{t} ( C^{SU}_t  y_{rp,k,t} + C^{UP}_t  u_{rp,k,t} +  C^{VAR}_t p_{rp,k,t} )  \nonumber \\
        +\sum_r C^{OM}_r p_{rp,k,r} +  \sum_s C^{OM}_s p_{rp,k,s}        + \sum_i C^{ENS} pns_{rp,k,i}
                                         \Big) \nonumber \\
   + \sum_{rp,k}  W^{RP}_{rp}  W^{K}_{k} \Big(   \sum_t ( C^{VAR}_t C^{RES+} res^+_{rp,k,t} + C^{VAR}_t C^{RES-} res^-_{rp,k,t} ) \nonumber  \\
                                               + \sum_s ( C^{OM}_s  C^{RES+} res^+_{rp,k,s} + C^{OM}_s  C^{RES-} res^-_{rp,k,s} )
                                         \Big) \nonumber \\
   + \sum_g C^{INV}_g x_g \label{eqn:ObjectiveFunction} \\
0 \leq pns_{rp,k,i} \leq D^P_{rp,k,i} \quad \forall rp,k,i \label{eqn:PNS} \\
x_g \in \mathbb{Z}^{+,0}, x_g \leq \overline{X}_g \quad \forall g \label{eqn:BoundInvestments}
\end{align}
\end{subequations}

Constraints \eqref{eqn:ThermalConstraints} contain all constraints regarding thermal generators: upward reserve requirement \eqref{eqn:TUpReserveRequ}; downward reserve requirement \eqref{eqn:TDnReserveRequ}; definition of total power output with the technical minimum and output above the technical minimum \eqref{eqn:TPowerOutput}; limit of upward reserve in case start-up occurred \eqref{eqn:TLimitReserveUp1}; limit of upward reserve in case shut-down occurs \eqref{eqn:TLimitReserveUp2}; limit of downward reserve \eqref{eqn:TLimitReserveDn}; definition of commitment, start-up and shut-down logic \eqref{eqn:TCommitmentLogic}; upper bound of commitment variable \eqref{eqn:TLimitCommitment}; ramp-up constraint \eqref{eqn:TRULimit}; ramp-down constraint \eqref{eqn:TRDLimit}; lower and upper bound of total power output \eqref{eqn:TUB1}; lower and upper bound of reserves and output above the minimum \eqref{eqn:TUB2}; definition of logical variables as binaries \eqref{eqn:TDefBinary}.
 
\begin{subequations}
\label{eqn:ThermalConstraints}
\begin{align}
\sum_t res^+_{rp,k,t} + \sum_s res^+_{rp,k,s} \geq RES^+ \sum_i D^P_{rp,k,i} \quad \forall rp,k   \label{eqn:TUpReserveRequ} \\
\sum_t res^-_{rp,k,t} + \sum_s res^-_{rp,k,s} \geq RES^- \sum_i D^P_{rp,k,i} \quad \forall rp,k   \label{eqn:TDnReserveRequ} \\
p_{rp,k,t} = u_{rp,k,t} \underline{P}_t + \hat{p}_{rp,k,t} \quad \forall rp,k,t  \label{eqn:TPowerOutput}\\
\hat{p}_{rp,k,t} + res^+_{rp,k,t} \leq (\overline{P}_t - \underline{P}_t) (u_{rp,k,t} - y_{rp,k,t}) \quad \forall rp,k,t \label{eqn:TLimitReserveUp1} \\
\hat{p}_{rp,k,t} + res^+_{rp,k,t} \leq (\overline{P}_t - \underline{P}_t) (u_{rp,k,t} - z_{rp,k++1,t}) \quad \forall rp,k,t \label{eqn:TLimitReserveUp2} \\
\hat{p}_{rp,k,t} \geq res^-_{rp,k,t}  \quad \forall rp,k,t \label{eqn:TLimitReserveDn} \\
u_{rp,k,t} - u_{rp,k--1,t} = y_{rp,k,t} - z_{rp,k,t} \quad \forall rp,k,t \label{eqn:TCommitmentLogic} \\
u_{rp,k,t} \leq x_{t} + EU_t \quad \forall rp,k,t \label{eqn:TLimitCommitment} \\
\hat{p}_{rp,k,t} - \hat{p}_{rp,k--1,t} + res^+_{rp,k,t} \leq u_{rp,k,t} RU_t \quad \forall rp,k,t \label{eqn:TRULimit} \\
\hat{p}_{rp,k,t} - \hat{p}_{rp,k--1,t} - res^-_{rp,k,t} \geq -u_{rp,k--1,t} RD_t \quad \forall rp,k,t \label{eqn:TRDLimit} \\
0 \leq p_{rp,k,t} \leq \overline{P}_t (x_t + EU_t) \quad \forall rp,k,t \label{eqn:TUB1} \\
0 \leq \hat{p}_{rp,k,t},res^-_{rp,k,t},res^+_{rp,k,t}  \leq (\overline{P}_t - \underline{P}_t) (x_t + EU_t) \quad \forall rp,k,t \label{eqn:TUB2} \\
u_{rp,k,t}, y_{rp,k,t},z_{rp,k,t}\in \{0,1 \} \quad \forall rp,k,t \label{eqn:TDefBinary}
\end{align}
\end{subequations}

Constraints \eqref{eqn:StorageConstraints} represent all standard constraints regarding storage technologies. Since we follow a flexible time representation methodology, as introduced in section \ref{subsec:ModelTime}, some of the constraints described in \eqref{eqn:StorageConstraints} only occur when we have representative days. In particular, when we use representative periods then we have two different types of storage state of charge constraints: intra-period (within the representative period); and inter-period (between different representative periods) constraints that are imposed on a moving window $MOW$ (for example once a week) throughout the time horizon. Note that MOW could cross multiple representative periods. If we have the exact hourly model, there is no need for inter-period constraints and all storage technologies would be modeled via the intra-period storage constraints. If we have representative periods then we require the inter-period storage constraint in order to model long-term effects that are important for hydro storage. An intra-period state of charge definition would not exist for a hydro storage technology. Since, this formulation is not novel, we refer the interested reader to \cite{Tejada-Arango18} where such a formulation is described in detail.

Constraint \eqref{eqn:SInterSOCDef} represents the inter-period evolution of the storage state of charge; 
upper bound of inter storage state of charge \eqref{eqn:SUBInterSOC}; lower bound of inter storage state of charge \eqref{eqn:SLBInterSOC}; 
cyclic storage constraint \eqref{eqn:SCyclicInter};
intra-period evolution of storage state of charge \eqref{eqn:SIntraSOCDef}; bound of upward reserve \eqref{eqn:SUBReserve}; bound of downward reserve \eqref{eqn:SLBReserve}; upper bound of intra storage state of charge \eqref{eqn:SUBIntraSOC}; lower bound of intra storage state of charge \eqref{eqn:SLBIntraSOC}; 
to avoid simultaneous charging and discharging \eqref{eqn:SAvoidChDis};
definition of binary variable to avoid simultaneous charging and discharging \eqref{eqn:SAvoidChDis2};
lower and upper bounds on production, consumption and reserve variables \eqref{eqn:SLBUB}; lower and upper bound of intra storage state of charge \eqref{eqn:SLBUBIntra}; 
lower and upper bound on spillages \eqref{eqn:SLBUBSpillage}.

\begin{subequations}
\label{eqn:StorageConstraints}
\begin{align}
inter_{p,s} = inter_{p-MOW,s} + InRes_{s,p=MOW} \nonumber \\
+ \sum_{\Gamma (p-MOW \leq pp\leq p,rp,k)} (- sp_{rp,k,s}
                                            + IF_{rp,k,s} W_k^K \nonumber \\
                                            - p_{ rp,k,s} W_k^K / \eta^{CH}_s 
                                            + cs_{rp,k,s} W_k^K \eta^{DIS}_s)
                                             \quad \forall p,s  
                                             \label{eqn:SInterSOCDef} \\
inter_{p,s} \leq \overline{P}_s ETP_s (x_{s} + EU_s)  \quad \forall s, p: mod(p,MOW)=0  \label{eqn:SUBInterSOC} \\
inter_{p,s} \geq \underline{R}_s \overline{P}_s ETP_s (x_{s} + EU_s)  \quad \forall s, p: mod(p,MOW)=0 \label{eqn:SLBInterSOC} \\
inter_{p,s} \geq InRes_{s,p}  \quad \forall s, p=CARD(p)  \label{eqn:SCyclicInter} \\
intra_{rp,k,s} = intra_{rp,k--1,s} - sp_{rp,k,s} + IF_{rp,k,s} W_k^K \nonumber \\
- p_{ rp,k,s} W_k^K / \eta^{CH}_s + cs_{rp,k,s} W_k^K \eta^{DIS}_s)
                                             \quad \forall rp,k 
                                             \label{eqn:SIntraSOCDef} \\
\hat{p}_{rp,k,s} - cs_{rp,k,s} + res^+_{rp,k,s} \leq \overline{P}_s (bx_{s} + EU_s) \quad \forall rp,k,s  \label{eqn:SUBReserve} \\
\hat{p}_{rp,k,s} - cs_{rp,k,s} - res^-_{rp,k,s} \geq -\overline{P}_s (bx_{s} + EU_s) \quad \forall rp,k,s \label{eqn:SLBReserve} \\
intra_{rp,k,s} \leq \overline{P}_s ETP_s (x_{s} + EU_s) \nonumber \\
- (res^-_{rp,k,s}+res^-_{rp,k--1,s}) W_k^K \quad \forall rp,k,s  \label{eqn:SUBIntraSOC}\\
intra_{rp,k,s} \geq \underline{R}_s \overline{P}_s ETP_s (x_{s} + EU_s) \nonumber \\
+ (res^+_{rp,k,s}+res^+_{rp,k--1,s}) W_k^K \quad \forall rp,k,s \label{eqn:SLBIntraSOC} \\
p_{ rp,k,s} \leq b^{ch/d}_{ rp,k,s} M^{ch/d},
cs_{ rp,k,s} \leq (1- b^{ch/d}_{ rp,k,s}) M^{ch/d}  \quad \forall rp,k,s \label{eqn:SAvoidChDis} \\
b^{ch/d}_{ rp,k,s} \in \{0,1 \}\quad \forall rp,k,s \label{eqn:SAvoidChDis2} \\
0 \leq p_{rp,k,s},cs_{rp,k,s}res^-_{rp,k,s},res^+_{rp,k,s} \leq \overline{P}_s (bx_s + EU_s) \quad \forall rp,k,s \label{eqn:SLBUB} \\
ETP_s \overline{P}_s \underline{R}_s (x_s + EU_s) \leq intra_{rp,k,s} \leq ETP_s \overline{P}_s (x_s + EU_s) \quad \forall rp,k,s \label{eqn:SLBUBIntra}\\
0 \leq sp_{rp,k,s} \leq (1-\overline{R}_s) ETP_s \overline{P}_s (x_s + EU_s)\quad \forall rp,k,s=hydro \label{eqn:SLBUBSpillage}
\end{align}
\end{subequations}



Constraints \eqref{eqn:RenewableConstraints} represent: lower and upper bounds on renewable production \eqref{eqn:RLBUB}; and, a system-wide constraint that limits thermal production to at most (1- $\kappa$) percent of total system demand \eqref{eqn:RClean}, thereby implicitly forcing $\kappa$ percent clean production.

\begin{subequations}
\label{eqn:RenewableConstraints}
\begin{align}
0 \leq p_{rp,k,r} \leq  \overline{P}_r  PF_{rp,k,r} (x_{r} + EU_r)
\quad \forall rp,k, r \label{eqn:RLBUB} \\
\sum_{rp,k,t} W_k^K W_{rp}^{RP} p_{rp,k,t} 
\leq (1-\kappa) \sum_{rp,k,i} W_k^K W_{rp}^{RP} D^P_{rp,k,i} \label{eqn:RClean}
\end{align}
\end{subequations}

Constraints \eqref{eqn:DC-OPFConstraints} represent the optimal power flow in DC: active power balance constraint \eqref{eqn:DC-Balance}; definition of power flow variable using injection shift factors \eqref{eqn:DC-Flow}; lower and upper bounds on power flow\eqref{eqn:DC-Bounds}.

\begin{subequations}
\label{eqn:DC-OPFConstraints}
\begin{align}
  \sum_{gi(t,i)} p_{rp,k,t} 
+ \sum_{gi(r,i)} p_{rp,k,i}
+ \sum_{gi(s,i)} (p_{rp,k,s} - cs_{rp,k,s}) \nonumber \\ 
+ \sum_{ijc(j,i,c)} f^P_{rp,k,j,i,c}
- \sum_{ijc(i,j,c)} f^P_{rp,k,i,j,c}
+ pns_{rp,k,i}                                 
=  D^P_{rp,k,i}  \quad \forall rp,k,i \label{eqn:DC-Balance} \\
f^P_{rp,k,ijc} = \sum_{iws} ISF(ijc,iws) 
(\sum_{gi(g,iws)} p_{rp,k,g} \nonumber \\
-\sum_{s,iws} cs_{rp,k,g}
+ pns_{rp,k,iws} -D^P_{rp,k,iws}) \quad \forall rp,k, ijc(i,j,c) \label{eqn:DC-Flow} \\
-\overline{T}_{i,j,c} \leq f^P_{rp,k,i,j,c} \leq \overline{T}_{i,j,c} \quad \forall rp,k,i,j,c  \label{eqn:DC-Bounds}
\end{align}
\end{subequations}

\subsection{Modeling Inertia}
\label{subsec:ModelInertia}

System inertia is a technical issue that is mainly discussed in the context of operational problems, but not in an expansion planning context. However, with global policy objectives of carbon-neutral power systems, omitting an important issues such as system inertia might lead to sub-optimal planning or even worse an infeasible technology mix. In \cite{Paturet2020}, the authors have derived a linear formulation of the RoCoF and Nadir constraints in an operational framework such as the unit commitment (UC) problem. In an extension \cite{paturet2020economic} of their work, in which they analyze pricing methodologies of inertia, they state that the RoCoF constraint is the main driver of inertia, and that the Nadir constraint could be omitted.

In this section, we extend the formulation presented in \cite{Paturet2020,paturet2020economic} to a generation expansion planning framework with UC operational constraints. Based on \cite{Paturet2020} we distinguish between two different sources of inertia: sources stemming from traditional synchronous generators; and, sources stemming from emulated/virtual inertia and based on power electronics. To that purpose we introduce index $v$ (with alias $vv$ and a subindex of $g$) representing all technologies that can provide virtual inertia, such as batteries or wind turbines for example. We make the assumption that inertia from synchronous generators and virtual inertia are equivalent in the context that one can be exchanged for the other.

For the sake of simplicity, we assume here that $\overline{X}_t=1$ whereas $\overline{X}_v>1$. In other words, we only allow for the investment of at most 1 thermal unit of each type\footnote{If one wanted to invest in two identical units, then this could be done easily by introducing another element with identical data to the set $t$. As a matter of fact, in our case study we have several almost identical thermal units.}. By doing so, we guarantee that $u_{rp,k,t}$ are binaries, as opposed to integer variables. Index $v$ on the other hand represents a type of virtual inertia providing technology, e.g., a wind farm, which has a standard size of 100 MW. The investment variable $x_v$ is an integer variable, whose value represents the multiple of the standard 100 MW wind farm. For example, if $x_v=2$ it would mean we are building a 200 MW wind farm. 

Let us now describe in detail the corresponding inertia constraints \eqref{eqn:InertiaConstraints} for a generation expansion framework:
definition of the scaled power gain factor of a thermal unit \eqref{eqn:M-Thermalk}, which represents how much one particular (dispatched) thermal unit can contribute with respect to the total dispatched thermal capacity in the system;
definition of the scaled power gain factor of a virtual unit (such as a wind turbine or a battery) \eqref{eqn:M-Virtualk}, which is defined slightly differently from the corresponding factor of a synchronous generator because it does not have a commitment variable. It is therefore defined as the fraction of its current power output over the total available virtual power output in this moment\footnote{In a system where there is a large amount of wind curtailment, this would be an overly conservative approximation.}.
Definition of inertia provided by synchronous generators \eqref{eqn:M-MSG} and virtual generators \eqref{eqn:M-MVI}. The right-hand side of \eqref{eqn:M-MVI} is multiplied by $x_v$ because $k_{rp,k,v}$ represents the power gain factor for one unit of the virtual technology $v$. However, in total $x_v$ total units of technology $v$ are built. In \eqref{eqn:M-MSG} this number is guaranteed to be 1, and that is why it is not explicitly modeled in \eqref{eqn:M-MSG}.
Definition of total system inertia \eqref{eqn:M-M}: note that total system inertia is a weighted average of virtual inertia and synchronous generator inertia.
Rate of change of frequency (RoCoF) constraint \eqref{eqn:M-ROCOF}.
Lower and upper bounds on inertia variables \eqref{eqn:M-LBUBM}.
Lower and upper bounds on power gain factors \eqref{eqn:M-LBUBk}.

\begin{subequations}
\label{eqn:InertiaConstraints}
\begin{align}
k_{rp,k,t} = \frac{\overline{P}_t}{\sum_{tt} \overline{P}_{tt} u_{rp,k,tt}} u_{rp,k,t} \quad \forall rp,k,t \label{eqn:M-Thermalk}\\
k_{rp,k,v} = \frac{p_{rp,k,t}}{\sum_{vv} \overline{P}_{vv} (x_{vv} + EU_{vv}) PF_{rp,k,vv} }  \quad \forall rp,k,v \label{eqn:M-Virtualk}\\
M^{SG}_{rp,k} = \sum_t 2 k_{rp,k,t} H_t \quad \forall rp,k \label{eqn:M-MSG}\\
M^{VI}_{rp,k} = \sum_v 2 k_{rp,k,v} H_v x_v  \quad \forall rp,k \label{eqn:M-MVI}\\
M_{rp,k} = \frac{M^{SG}_{rp,k} \sum_{tt} \overline{P}_{tt} u_{rp,k,tt} + M^{VI}_{rp,k}\sum_{vv} \overline{P}_{vv} (x_{vv} + EU_{vv}) PF_{rp,k,vv}}{\sum_{tt} \overline{P}_{tt} u_{rp,k,tt} + \sum_{vv} \overline{P}_{vv} (x_{vv} + EU_{vv}) PF_{rp,k,vv}} \nonumber \\
\quad \forall rp,k \label{eqn:M-M}\\
\frac{\dot f_{lim}}{f_b} M_{rp,k} \geq \Delta P_{rp,k} \quad \forall rp,k \label{eqn:M-ROCOF}\\
0 \leq M^{VI}_{rp,k},M^{SG}_{rp,k},M_{rp,k} \leq \overline{M}  \quad \forall rp,k \label{eqn:M-LBUBM}\\
0 \leq k_{rp,k,g} \leq 1  \quad \forall rp,k,g \label{eqn:M-LBUBk}
\end{align}
\end{subequations}

Some of the constraints in \eqref{eqn:InertiaConstraints}, and in particular constraints \eqref{eqn:M-Thermalk}, \eqref{eqn:M-Virtualk}, \eqref{eqn:M-MVI}, and \eqref{eqn:M-M}, are nonlinear and would therefore complicate the linear nature of the other constraints of this investment model. Therefore, we proceed by linearizing the previously mentioned constraints. Note that since all of the non-linearities represent bilinear terms that are a product of a continuous and a discrete variable, the linearization is exact.

As a demonstration, we show how we linearize constraint \eqref{eqn:M-Thermalk}. For the sake of brevity, we do not explicitly show the linearization of all the remaining terms in this paper because they are all very similar to the linearization of \eqref{eqn:M-Thermalk}.

Constraint \eqref{eqn:M-Thermalk} is nonlinear because there is a variable, i.e., the commitment decision, in the denominator of the right-hand side. If we were to multiply both sides by this denominator\footnote{Assuming that the denominator is not zero. In the optimization model, there is an additional constraint that ensures $k_{rp,k,t}$ to be zero if the denominator is zero as well.}, then we obtain:

\begin{equation*}
  k_{rp,k,t} \sum_{tt} \overline{P}_{tt} u_{rp,k,tt} =  \overline{P}_t u_{rp,k,t} \quad \forall rp,k,t  
\end{equation*}

The right-hand side of this expression is linear, and on the left-hand side we obtain the product of the continuous variable $k_{rp,k,t}$ with the binary variable $u_{rp,k,tt}$. Let us now define an auxiliary continuous variable $ku^{aux}_{rp,k,tt,t}$ that should represent the product of these two variables. The following constraints assign the correct meaning to this variable, and therefore nonlinear constraint \eqref{eqn:M-Thermalk} is replaced by its linear equivalent \eqref{eqn:LinearizeIntertiak}.

\begin{subequations}
\label{eqn:LinearizeIntertiak}
\begin{align}
 \sum_{tt} \overline{P}_{tt} ku^{aux}_{rp,k,tt,t} =  \overline{P}_t u_{rp,k,t} \quad \forall rp,k,t  \\
 0 \leq ku^{aux}_{rp,k,tt,t} \leq u_{rp,k,tt} \quad \forall rp,k,tt,t  \\
 ku^{aux}_{rp,k,tt,t} \leq k_{rp,k,t} \leq 1 \quad \forall rp,k,tt,t  \\
  k_{rp,k,t} - ku^{aux}_{rp,k,tt,t} \leq 1 - u_{rp,k,tt} \quad \forall rp,k,tt,t  
\end{align}
\end{subequations}

Let us briefly discuss the remaining nonlinear constraints. Constraint \eqref{eqn:M-Virtualk} is also nonlinear because of the variables, i.e., the investment variable $x_{vv}$, in the denominator. We can also multiply both sides with the denominator and obtain a linear right-hand side, and a nonlinear left-hand side, which is the product of the continuous $k_{rp,k,v}$ with the discrete $x_{vv}$. Note that a discrete variable can be written as the sum of binary variables, which would render the bilinear terms the product of a continuous with a binary variable. In \eqref{eqn:LinearizeIntertiak} we have shown how to linearize such terms. We do the same here. 
In the remaining nonlinear constraints \eqref{eqn:M-MVI} and \eqref{eqn:M-M}, we find the same: products of continuous variables (either $k$, $M$, $M^{VI}$ or $M^{SG}$) with discrete or binary variables ($u$ or $x$). Again, we do not include all these linearizations here because they are tedious and in no way different from \eqref{eqn:LinearizeIntertiak}.

Therefore, please note that in the remainder of this paper, when we refer to inertia constraints \eqref{eqn:InertiaConstraints}, we really mean their linear equivalent.

\subsection{Modeling Reactive Power}
\label{subsec:ModelReactivePower}

In this section we introduce the constraints in order to replace the standard DC-OPF formulation with the full AC version \cite{Tang2015}. In particular, we employ the relaxation of the full AC-OPF via the second order cone programming (SOCP) formulation as in \cite{Low2014a} and \cite{Low2014b}. This formulation allows us to explicitly introduce variables such as reactive power, and voltages in the generation expansion model, as opposed to relying on the simplified DC-OPF.

From a mathematical point of view, considering the AC-OPF (as opposed to the DC-OPF) in the generation expansion model means replacing constraints \eqref{eqn:DC-OPFConstraints} with \eqref{eqn:AC-OPFConstraints}. We also introduce new auxiliary variables that are used in the SOCP formulation: $cii_{rp,k,i}, cij_{rp,k,i,j}, sij_{rp,k,i,m} $, which represent the square of the voltage at bus $i$, the product of the voltages at bus$i$ and $j$ times $cos(\theta_{ij})$, and the product of the voltages at bus$i$ and $j$ times $sin(\theta_{ij})$.

Constraints \eqref{eqn:AC-OPFConstraints} represent:
an updated version\footnote{Note that the active power balance constraint used in the SOCP formulation is not the same as in the DC-OPF. The SOCP formulation explicitly models a term that uses the square of the bus voltages, which causes differences between the DC and the SOCP results.} of the active power balance equation \eqref{eqn:AC-BalanceP};
a reactive power balance \eqref{eqn:AC-BalanceQ};
the conic constraint of the SOCP representing the squares of voltages \eqref{eqn:AC-SOCP};
ensuring maximum angle differences \eqref{eqn:AC-AngleDif1};
definition of active power flow from bus $i$ to bus $j$ \eqref{eqn:AC-FlowP1};
definition of active power flow from bus $j$ to bus $i$ \footnote{In the DC-OPF we do not explicitly consider both directions of the power flow, demonstrated by the fact that equation \eqref{eqn:DC-Flow} is only defined in one direction. This is connected to the underlying DC hypothesis that $f^P_{i,j}=-f^P_{i,j}$, which is not necessarily the case in actual AC power flow. That is why in the SOCP formulation we have to explicitly define and distinguish between the direction of the power flow.} \eqref{eqn:AC-FlowP2};
definition of reactive power flow from bus $i$ to bus $j$ \eqref{eqn:AC-FlowQ1};
definition of reactive power flow from bus $j$ to bus $i$ \eqref{eqn:AC-FlowQ2};
lower and upper bounds on reactive power provided by FACTS \eqref{eqn:AC-LBUBqFACTS};
lower and upper bounds on reactive power provided by thermal units \eqref{eqn:AC-LBUBqThermal};
lower and upper bounds on reactive power provided by unit $g$ \eqref{eqn:AC-LBUBq}\footnote{This constraint is redundant for thermal and FACTS given that there are more binding constraints for them available. This is just a general limit to help the numerical solvers, and for storage and renewable units should they be able to provide reactive power.};
bounds on active power flow \eqref{eqn:AC-LBUBfp};
bounds on reactive power flow \eqref{eqn:AC-LBUBfq};
bounds of auxiliary $cii$ variable \eqref{eqn:AC-LBUBcii};
bounds of auxiliary $cij$ variable \eqref{eqn:AC-LBUBcij};
bounds of auxiliary $sij$ variable \eqref{eqn:AC-LBUBsij}.

\begin{subequations}
\label{eqn:AC-OPFConstraints}
\begin{align}
  \sum_{gi(t,i)} p_{rp,k,t} 
+ \sum_{gi(r,i)} p_{rp,k,i}
+ \sum_{gi(s,i)} p_{rp,k,s}
- \sum_{gi(s,i)} cs_{rp,k,s}  
+ pns_{rp,k,i}                                  \nonumber \\ 
=
  \sum_{(j,c) \in ijc(i,j,c)} f^P_{rp,k,i,j,c}
+ \sum_{(j,c) \in ijc(j,i,c)} f^P_{rp,k,i,j,c}  \nonumber \\
+ cii_{rp,k,i} G_i SB                       
+ D^P_{rp,k,i}  
\quad \forall rp,k,i \label{eqn:AC-BalanceP} \\
  \sum_{gi(t,i)} q_{rp,k,t} 
+ \sum_{gi(r,i)} q_{rp,k,i}
+ \sum_{gi(s,i)} q_{rp,k,s}
+ \sum_{gi(facts,i)} q_{rp,k,facts}   \nonumber \\
+ pns_{rp,k,i} R_i                             
=
  \sum_{(j,c) \in ijc(i,j,c)} f^Q_{rp,k,i,j,c}
+ \sum_{(j,c) \in ijc(j,i,c)} f^Q_{rp,k,i,j,c}  \nonumber \\
- cii_{rp,k,i} B_i SB                       
+ D^Q_{rp,k,i}  
\quad \forall rp,k,i  \label{eqn:AC-BalanceQ} \\
cij^{2}_{rp,k,i,j} + sij^{2}_{rp,k,i,j} \leq cii_{rp,k,i} cii_{rp,k,j} \quad \forall rp,k,line(i,j)  \label{eqn:AC-SOCP}\\
-cij_{rp,k,i,j} tan(\Delta) \leq sij_{rp,k,i,j} \leq  cij_{rp,k,i,j} tan(\Delta) \quad \forall rp,k,line(i,j) \label{eqn:AC-AngleDif1} \\
f^P_{rp,k,i,j,c} = SB [ G_{i,j,c} cii_{rp,k,i} - cij_{rp,k,i,j} G_{i,j,c} + sij_{rp,k,i,j} B_{i,j,c}] \nonumber \\
                      \quad \forall rp,k,ijc(i,j,c) \label{eqn:AC-FlowP1} \\
f^P_{rp,k,j,i,c} = SB [ G_{i,j,c} cii_{rp,k,j} - cij_{rp,k,i,j} G_{i,j,c} - sij_{rp,k,i,j} B_{i,j,c}] \nonumber \\
                      \quad \forall rp,k,ijc(i,j,c) \label{eqn:AC-FlowP2} \\
f^Q_{rp,k,i,j,c} = SB [- (B_{i,j,c}+Bc_{i,j,c}/2) cii_{rp,k,i} + sij_{rp,k,i,j} G_{i,j,c} + cij_{rp,k,i,j} B_{i,j,c}] \nonumber \\
                      \quad \forall rp,k,ijc(i,j,c) \label{eqn:AC-FlowQ1} \\
f^Q_{rp,k,j,i,c} = SB [- (B_{i,j,c}+Bc_{i,j,c}/2) cii_{rp,k,j} - sij_{rp,k,i,j} G_{i,j,c} + cij_{rp,k,i,j} B_{i,j,c}] \nonumber \\
                      \quad \forall rp,k,ijc(i,j,c) \label{eqn:AC-FlowQ2} \\
x_{facts} \underline{Q}_{facts} \leq q_{rp,k,facts} \leq x_{facts} \overline{Q}_{facts} \quad \forall rp,k,facts \label{eqn:AC-LBUBqFACTS} \\
u_{rp,k,t} \underline{Q}_t \leq q_{rp,k,t} \leq u_{rp,k,t} \overline{Q}_t \quad \forall rp,k,t \label{eqn:AC-LBUBqThermal} \\
\underline{Q}_{g} \leq q_{rp,k,g} \leq \overline{Q}_{g} \quad \forall rp,k,g \label{eqn:AC-LBUBq} \\
-\overline{T}_{i,j,c} \leq f^P_{rp,k,i,j,c} \leq \overline{T}_{i,j,c} \quad \forall rp,k,ijc(i,j,c)  \label{eqn:AC-LBUBfp} \\
- \overline{A}_{i,j,c} \leq f^Q_{rp,k,i,j,c} \leq \overline{A}_{i,j,c} \quad \forall rp,k,ijc(i,j,c) \label{eqn:AC-LBUBfq} \\
\underline{V}_{i}^2 \leq cii_{rp,k,i} \leq \overline{V}_{i}^2 \quad \forall rp,k,i \label{eqn:AC-LBUBcii} \\
\underline{V}_{i}^2 \leq cij_{rp,k,i,j} \leq \overline{V}_{i}^2 \quad \forall rp,k,line(i,j) \label{eqn:AC-LBUBcij} \\
- \overline{V}_{i}^2 \leq sij_{rp,k,i,j} \leq \overline{V}_{i}^2 \quad \forall rp,k,line(i,j) \label{eqn:AC-LBUBsij} 
\end{align}
\end{subequations}

\subsection{Overview of Model Options}
\label{subsec:OverviewModels}
Since we have introduced many equations in section \ref{sec:Methodology}, let us provide the reader with an overview of the different types of case studies that one can carry out assembling different LEGO blocks. In particular, Table \ref{tab:OverviewModels} gives a summary of the different cases (and their corresponding constraints) that we discuss in this paper. First, in this paper we only solve generation expansion models; however, if one wanted to run an operation-only case, this could be achieved easily by simply fixing the investment variables.

Second, all of the generation expansion model cases we run here contain: the standard constraints of the objective function and other general constraints \eqref{eqn:GeneralConstraints}; unit commitment constraints for traditional thermal units \eqref{eqn:ThermalConstraints}; constraints for storage technologies \eqref{eqn:StorageConstraints}; and, renewable generators and clean production constraints \eqref{eqn:RenewableConstraints}. What differentiates the model cases is how they treat: the network representation (DC \eqref{eqn:DC-OPFConstraints} or AC \eqref{eqn:AC-OPFConstraints}), and inertia \eqref{eqn:InertiaConstraints}. 

The most basic generation expansion model that we can build out of our LEGO blocks does not consider inertia constraints, and uses a DC approximation of the power flow. We refer to this as the Base Case (BC), which represents the simplistic version. The other, most realistic, version would be the full LEGO model, which does factor in both inertia constraints, and considers power flow equations in AC (as approximated by the SOCP).
The remaining cases represent a sensitivity analysis regarding the BC, where we: 
simply add the inertia block to the BC and obtain what we refer to as the Inertia Case (IC); or, exchange the DC-block \eqref{eqn:DC-OPFConstraints} with the AC-block \eqref{eqn:AC-OPFConstraints} and obtain the Reactive Case (RC) study. Both RC, and IC, can be viewed as intermediate steps between the BC and the full LEGO model, which allow us to analyze the separate impacts that inertia and reactive power constraints may have on GEP.

\begin{table}[ht]
    \centering
    \begin{tabular}{c|c|c}
    \textbf{Model Name} &\textbf{Model Description} & \textbf{Constraints} \\ \hline 
    Base Case           & GEP + UC + DC-OPF            &  \eqref{eqn:GeneralConstraints},\eqref{eqn:ThermalConstraints},\eqref{eqn:StorageConstraints},\eqref{eqn:RenewableConstraints},\eqref{eqn:DC-OPFConstraints}      \\
    (BC)                & no inertia        &           \\  \hline    
    Reactive Case       & GEP + UC + \textbf{AC-OPF}           &  \eqref{eqn:GeneralConstraints},\eqref{eqn:ThermalConstraints},\eqref{eqn:StorageConstraints},\eqref{eqn:RenewableConstraints}     \\
    (RC)                & no inertia        &  \eqref{eqn:AC-OPFConstraints}         \\  \hline 
    Inertia Case        & GEP + UC + DC-OPF              & \eqref{eqn:GeneralConstraints},\eqref{eqn:ThermalConstraints},\eqref{eqn:StorageConstraints},\eqref{eqn:RenewableConstraints},\eqref{eqn:DC-OPFConstraints}       \\
    (IC)                & + \textbf{inertia}        &  \eqref{eqn:InertiaConstraints}           \\  \hline
    Full LEGO Model      & GEP + UC + \textbf{AC-OPF}              & \eqref{eqn:GeneralConstraints},\eqref{eqn:ThermalConstraints},\eqref{eqn:StorageConstraints},\eqref{eqn:RenewableConstraints},       \\
    (LEGO)                & + \textbf{inertia}            & \eqref{eqn:InertiaConstraints},\eqref{eqn:AC-OPFConstraints}           \\  \hline
    \end{tabular}
    \caption{Models options and corresponding constraints}
    \label{tab:OverviewModels}
\end{table}

In the different case studies discussed in the following sections, we sometimes refer to different percentages of clean (carbon-free) production. These different percentages do not refer to different models. They simply refer to having changed the value of parameter $\kappa$ in one of the standard constraints \eqref{eqn:RClean}. Note that $\kappa=0$ practically relaxes the constraint.

\section{Numerical Results}
\label{sec:Results}

Section \ref{sec:Results} starts out by briefly discussing in \ref{subsec:Data} the data  used in the numerical studies that follow. 
In section \ref{subsec:ResultsInertia}, we analyze the impact of the novel inertia constraints in GEP (BC versus IC). Section \ref{subsec:ResultsReactive} shows how including reactive power constraints in GEP can change traditional base case results (BC versus RC). Finally, in section \ref{subsec:ResultsHolistic} we present the results of the full LEGO model that factors in both inertia and reactive power constraints. 

\subsection{Data}
\label{subsec:Data}

The case study analyzed in this paper covers a time horizon of one static year in the future, which has been approximated by 7 representative days. The complete data set used in this paper is available online\footnote{https://github.com/wogrin/LEGO.git} and is based on the StarNet Lite demo version\footnote{https://www.iit.comillas.edu/aramos/starnet.htm} for long-term planning developed by Prof. Andres Ramos at IIT-Comillas. 
We have extended the original data set to include storage from \citep{TejadaArango2020} and renewable generation and corresponding profiles from \cite{PFENNINGER20161251} and \cite{STAFFELL20161224}, and information about inertia, reactive power, FACTS etc from \cite{Kundur1994}. Even though the detailed data set is available online, we briefly outline the main features of the case to provide the reader with an overview.

\begin{figure}[h]
\includegraphics[width=\textwidth]{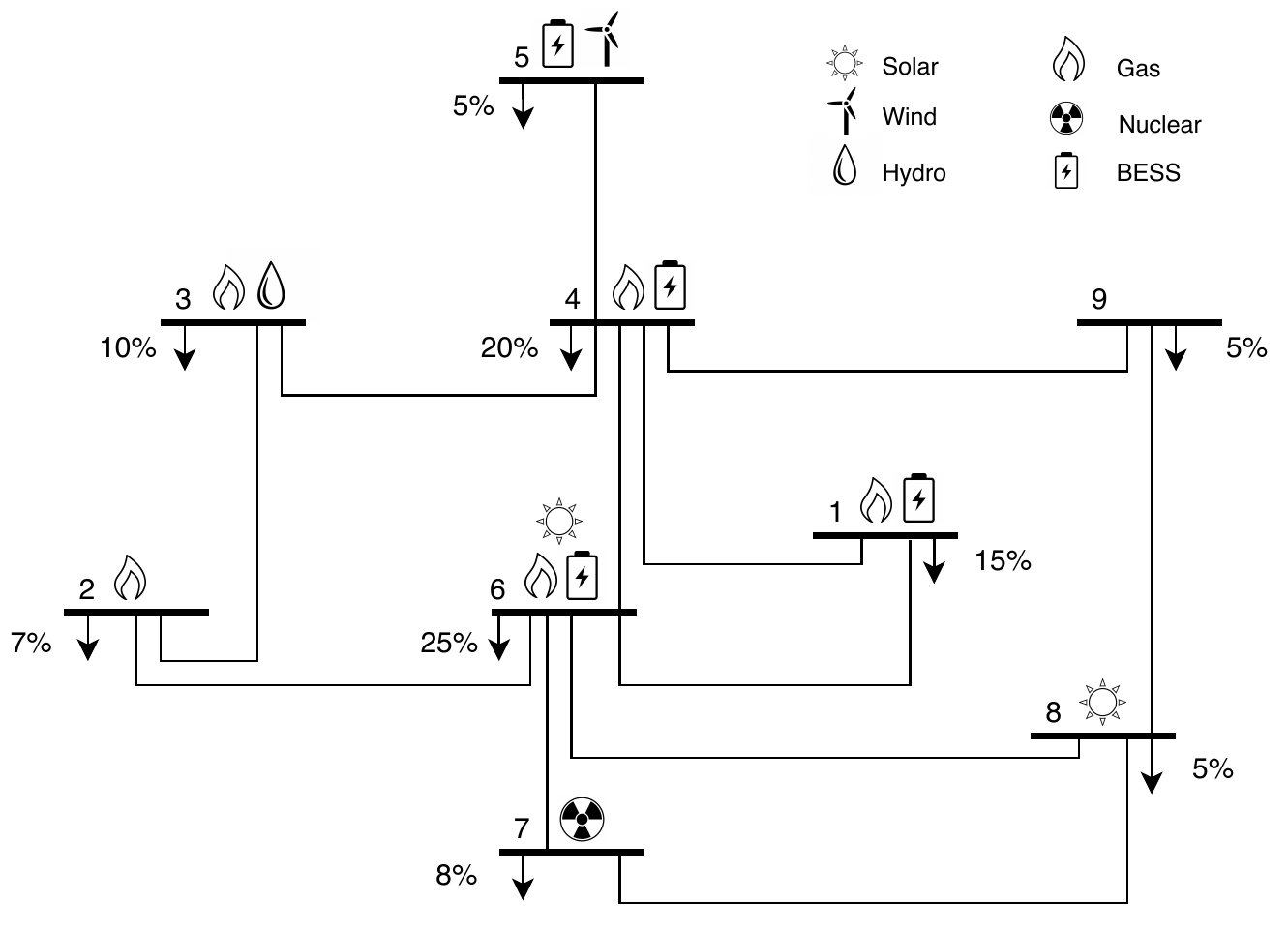}
\caption{9-bus network, existing (nuclear and hydro) and candidate (everything else) generation units and nodal demand indicated in percent.}
\label{fig:Network}
\end{figure}

We consider a 9-bus network with 13 existing transmission lines with an 800MVA capacity limit on each line and a $\pm$ 10\% voltage limit for each node. The network itself, as well as the location of the existing and candidate generating units are depicted in Figure \ref{fig:Network}. We indicate the percent of total demand consumed at each node. At each of the 9 buses, there is the option of installing a FACTS device for reactive power management. Since in our results neither coal nor fueloil units are ever built, we omit them from the diagram in Figure \ref{fig:Network}.

The most important data for thermal generator types is given in Table \ref{tab:Data-ThermalGen}.
The only existing generators are one 772 MW nuclear unit located at bus 7, and one 600 MW pumped hydro plant at bus 3. As for candidate thermal generation, we consider four different coal units located at nodes 1 to 4, four combined cycle gas turbine (CCGT) units at buses (1,3,4 and 6)\footnote{$CCGT^*$ is located at bus 1. Units $CCGT^{**}$ are located at buses 3,4 and 6.}, three open cycle gas turbines (OCGT) at nodes 2, 4 and 6, and one fueloil unit at bus 9\footnote{Coal and fueloil units are omitted in Figure \ref{fig:Network} because the model never decides to build them.}. The upper bound on investments per unit and per bus for all thermal technologies is one; however, there exist multiple CCGT and OCGT units within the network to counterbalance this limit. 

\begin{table}[ht]
    \centering
    \begin{tabular}{l|c|c|c|c|c|c|c}
                                & $\underline{P} ; \overline{P}$& $\underline{Q} ; \overline{Q}$& $H$ & $C^{SU}$   & $C^{VAR}$  & $C^{UP}$  & $C^{INV}$  \\  
                                & (MW)                          & (MVar)                        & (s) & (M\euro)   & (\euro/MWh)& (M\euro/h)& (M\euro/GW/year)\\  \hline 
    \textbf{Nuclear}            & 772 ; 772                 & 0 ; 0                         & 8   & -          & 15         & -         & -   \\  
    \textbf{CCGT$^{*}$}         & 134 ; 668                     & $\pm$ 200                     & 4   & 0.03       & 28         & 0.009     & 45.5   \\ 
    \textbf{CCGT$^{**}$}        & 100 ; 500                     & $\pm$ 267                     & 4   & 0.03       & 39         & 0.009     & 20.1   \\ 
    \textbf{OCGT }              & 40 ; 400                      & $\pm$ 180                     & 2.5 & 0.06       & 64         & 0.003     & 9.9   \\  \hline
    \end{tabular}
    \caption{Data for thermal generator types}
    \label{tab:Data-ThermalGen}
\end{table}

As for renewable technologies, we consider wind and solar and present the most important data per generator type in Table \ref{tab:Data-RenewableGen}. Note that the model itself can build an unlimited amount of each generator type. The only constraint is the candidate location that is imposed by the network given in Figure \ref{fig:Network}. For wind, we consider two different investment options: a traditional one; and, one that can provide virtual inertia (VI) to the system. The latter investment option is slightly more expensive than the traditional one, and also has a higher operations and maintenance costs. The only location where any of these wind technologies can be built is bus 5. We set up the test case this way to represent many real power systems, e.g. Texas, where wind is far away from demand centers. Such a setup also allows for the interpretation of having an off-shore wind farm for example. As for solar, we consider two potential locations at bus 6 and 8 respectively. Bus 6 is the main demand center, and solar investment there can be interpreted as building rooftop panels, whereas bus 8 can be seen as a remote location with very little demand, where a large solar farm can be built. Both wind and solar technologies dispose of an hourly maximum production profile (sometimes referred to as availability factor).

\begin{table}[ht]
    \centering
    \begin{tabular}{l|c|c|c|c}
                      & $\overline{P}$ & $H$ & $C^{OM}$   & $C^{INV}$  \\  
                      & (MW)           & (s) & (\euro/MWh)& (M\euro/GW/year)\\  \hline 
    \textbf{Wind}     & 100            & 0   & 2          & 7.3  \\  
    \textbf{Wind VI}  & 100            & 2   & 5          & 8  \\ 
    \textbf{Solar }   & 100            & 0   & 0          & 8.4  \\  \hline
    \end{tabular}
    \caption{Data for renewable generator types}
    \label{tab:Data-RenewableGen}
\end{table}

Within the technology option of battery energy storage systems (BESS) we differentiate between two types: a traditional one; and, one that provides virtual inertia. The most important data is presented in Table \ref{tab:Data-StorageGen}. Note that the data corresponds to one BESS unit; however, the model is not constrained with respect to the total amount of units, just with respect to the location of these units. Both operation and maintenance costs and investment costs are higher for the VI batteries. The candidate nodes for BESS investments (of both types) are nodes 1, 4, 5, and 6. The inertia constant $H_v$ on how much VI could be provided by a BESS and a wind unit has been taken from \cite{phdthesisUros}.

\begin{table}[ht]
    \centering
    \begin{tabular}{l|c|c|c|c|c|c}
                      & $\overline{P}$ & $H$ & $C^{OM}$   & $C^{INV}$        & $\eta^{CH/DIS}$ & ETP \\  
                      & (MW)           & (s) & (\euro/MWh)& (M\euro/GW/year) & (\%)            & (h)\\  \hline 
    \textbf{BESS}     & 100            & 0   & 4          & 3.2              & 95              & 4\\  
    \textbf{BESS VI}  & 100            & 10  & 10         & 3.4              & 95           & 4\\  \hline
    \end{tabular}
    \caption{Data for storage generator types}
    \label{tab:Data-StorageGen}
\end{table}

In this case study we consider that only BESS VI and Wind VI can provide virtual inertia. We do not consider a separate solar technology that provides virtual inertia, even though theoretically this could be done. However, the most realistic way for a solar PV generator to provide emulated inertia would be to couple it with a battery unit, and since we do allow for investments in virtual-inertia-providing batteries in the case study we do not introduce a separate Solar VI technology. Wind turbines on the other hand could provide inertia without having to be coupled with a battery. If the controller were to be adapted to account for changes in frequency, wind turbines can make use of the kinetic energy of the rotating mass to provide inertia. 

Within our numerical results, we study different cases of policy goals. In particular, we limit the percentage of thermal production with respect to total system demand as represented by constraint \eqref{eqn:RClean}. In particular, the parameter $\kappa$, which we sometimes refer to as renewable penetration, takes values 33\% (the business as usual case), 50\%, 70\%, 90\% and 100\% (no thermal production allowed). Values of $\kappa$ lower than 33\% would yield the exact the same results as the 33\% case itself, as 33\% is the renewable penetration that arises naturally (without imposing any policy goals) due to the operating and investment costs that have been assumed. We sometimes refer to the 100\%-case as the \textit{carbon-neutral} case, however, we want to specify that - since we are limiting thermal production - we are also excluding nuclear production here, which could also be considered carbon-neutral. 

All models have been run on an Intel Xeon with 2.60GHz and 144 GB RAM, were implemented using GAMS and solved using Gurobi 9.0.2. The model sizes range from 33148 constraints and 28871 variables (8083 integer) for the simplest BC model, to 401242 constraints and 162164 variables (8152 integer) for the full LEGO model. The corresponding CPU times are heavily case dependent and range from 26 seconds to 10 hours.
There are multiple reasons for this: for example, the inertia constraints \eqref{eqn:InertiaConstraints} require man linearizations of bilinear terms introducing a large amount of discrete variables into the model, which cause the branch and bound algorithms to take longer. Therefore, models that explicitly consider inertia constraints, take longer to solve. Moreover, $\kappa$ seems to have a high influence on run times as well. This is due to the fact that when $\kappa$ is high, i.e., a high renewable penetration, traditional thermal generators are used less (or not at all), which renders all corresponding unit commitment constraints superfluous. This also means that discrete investment decisions are predominantly renewable or BESS - all of which increases the amount of binary variables that have to be evaluated. On the other hand, if $\kappa$ is low then the model has to factor in UC constraints, which involve many discrete variables and, the model has to consider both renewable, BESS and thermal technologies in investment decisions, which ultimately leads ot higher CPU times.

\subsection{Sensitivity regarding Inertia Constraints}
\label{subsec:ResultsInertia}

As pointed out in data section \ref{subsec:Data}, in this power system we consider that both wind and battery technologies can provide inertia. Hence, in all the results we specifically differentiate those different technologies as, e.g. Wind and Wind VI (virtual inertia).


First of all, as the BC GEP model does not consider inertia, it does not build any virtual inertia providing plants (such as BESS VI or Wind VI), instead it builds their traditional equivalent (BESS or Wind). Therefore, the BC optimal generation mix determined by BC runs into problems when checking for inertia criteria ex-post. That is, during many time periods (when renewables are providing the large majority of demand) system inertia is low or even zero, violating the RoCoF constraint. 
In order to fix this problem should it arise, we can allow for changes in operating decisions ex-post. Computationally speaking: we run the BC model; then, we fix the investment variables, but we allow for generating units to alter dispatch and operating decisions; then, we run the IC model (but with fixed BC investments). The difference in the objective function between the BC and the constrained IC gives us an idea about the additional cost of making sub-optimal BC investment decisions inertia feasible. 

However, when running the constrained IC model enforcing that RoCoF is met for the predetermined BC expansion plan, operations are changed and traditional synchronous machines - the only ones that can provide inertia under the current mix - are dispatched more often, thereby violating the clean production constraint. Or in other words, the BC capacity mix either does not provide sufficient inertia or if it does, then it does not satisfy the policy goal. 

\begin{figure}[h]
\includegraphics[width=\textwidth]{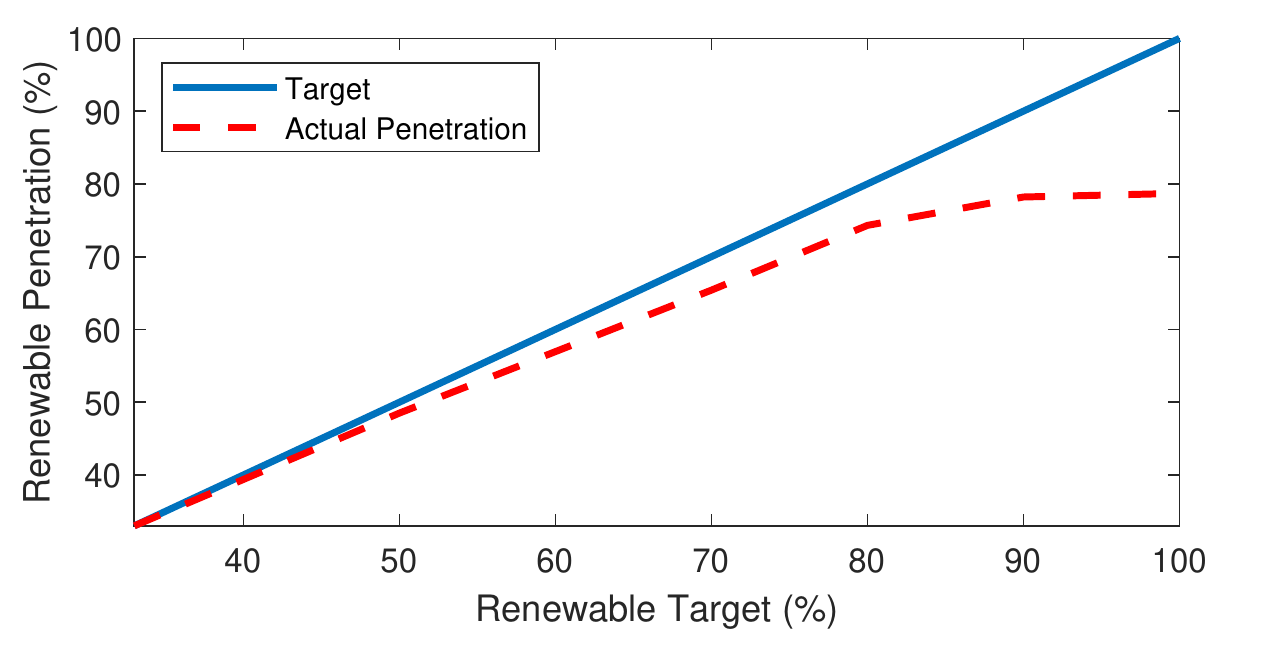}
\caption{Renewable penetration target versus actual renewable penetration with BC capacity mix and imposing inertia constraints ex-post.}
\label{fig:Target}
\end{figure}

In the observed case studies only the business as usual case capacity mix was able to provide inertia, by changing operating decisions ex-post, and also to comply with the policy goal. From 50\% clean production goal onward, doing both is no longer possible. If we disregard the renewable target, however, the additional system cost of satisfying inertia constraints ranges from 0.5 to 3.1\% of total system cost as can be seen in Table \ref{tab:Results-OperationInertia}. Regarding the deviation from the target, the optimal BC mix with a 50\% renewable goal, only achieves an actual of 48.4\% of renewable penetration when having to comply with RoCoF constraints. In Figure \ref{fig:Target} we observe how the higher the renewable target, the further the BC optimal mix deviates from this target when having to impose RoCoF constraints in operations ex-post. As a matter of fact, the results indicate that an implicit threshold of 78.7\% of actual renewable penetration, even when the target was a 100\% renewable penetration. Even though no new thermal plants are built, the existing nuclear plant is operated and provides 22.3\% of total system demand always, just to meet system inertia requirements. If that nuclear plant had not been in the existing capacity mix, the situation would be even worse - the system mix would actually be infeasible and physically incapable of providing inertia. Such a power system would simply not work. The message of this study is that when completely disregarding inertia in GEP, a renewable penetration beyond a certain point - 80\% renewable penetration in this case study - is simply impossible.

\begin{table}[ht]
    \centering
    \begin{tabular}{l|c|c|c|c|c}
    \textbf{Clean Prod. (\%)}          & 33         & 50         & 70         & 90         & 100       \\   \hline 
    \textbf{Total Cost BC }            & 1260.9     & 1315.1     & 1527.9     & 1933.5     & 2525.5   \\  
    \textbf{Actual Cost BC}            & (1260.9)   & (1320.9)   & (1546.7)   & (1982.1)   & (2605.5)    \\ 
    \textbf{Actual Clean \%}           & (33)       & (48.4)     & (65.4)     & (78.2)     & (78.7)    \\ \hline
    \end{tabular}
    \caption{Total system cost and renewable penetration under BC planning (allowing for changes in operation only when imposing inertia constraints ex-post) }
    \label{tab:Results-OperationInertia}
\end{table}

In other words, without the provision of virtual inertia by clean technologies, such as Wind VI or BESS VI for example, a power system cannot exceed a certain threshold of renewable penetration while at the same time guaranteeing RoCoF system stability. This is not surprising because the system has less and less thermal capacity available, which is operating more and more to satisfy inertia constraints, but there is a certain limit of clean production versus the total amount of inertia provisions that thermal plants can provide. Moreover, if traditional thermal plants are being forced out of the power system completely due to policy objectives, system inertia is decreasing continuously to a point where either the system cannot function any longer, because of the lack of synchronous generators, or the renewable targets are not met. This is an important message, because in terms of policy goals many countries are on their way to a carbon-neutral power system; however, virtual inertia providing technologies have not been introduced in power systems at a large scale yet. In order to achieve ambitious policy goals, substantial research and testing of how to build and integrate such technologies is urgently necessary.

Let us now continue this thought experiment and allow for additional investments to be made that do account for RoCoF constraints. In terms of model runs this would mean: run the BC model; fix the lower bound on investments; and re-run the IC model that explicitly captures inertia constraints. By allowing additional investments in technologies that provide virtual inertia (Wind VI and BESS VI in our case study), we can make the system RoCoF feasible at a relatively small additional cost which is given in parentheses in Table \ref{tab:Results-InvestmentInertia}. Note that additional capacity investment decisions are also presented in parentheses.

\begin{table}[ht]
    \centering
    \begin{tabular}{l|c|c|c|c|c}
    \textbf{Clean Prod. (\%)}          & 33         & 50         & 70         & 90         & 100       \\   \hline 
    \textbf{Total Cost BC }            & 1260.9     & 1315.1     & 1527.9     & 1933.5     & 2525.5   \\  
    \textbf{Actual Cost BC}            & (1260.9)   & (1324.9)   & (1542.5)   & (1959.1)   & (2560.0)    \\ 
    \textbf{Total Cost IC}             & [1260.9]   & [1321.7]   & [1541.0]   & [1946.5]   & [2539.9]    \\ 
    \hline
    \textbf{CCGT (GW)}                 &  1.7       & 1.7        & 0.7        & 0          & 0            \\  \hline
    \textbf{OCGT (GW)}                 &  0.8       & 0.4        & 0.4        & 0.4        & 0            \\  \hline
    \textbf{Wind (GW)}                 &  2.5       & 2.5        & 2.5        & 2.3        & 3.5 [2.3]     \\ \hline
    \textbf{Wind VI (GW)}              &  0         & 0          & 0          & 0          & 0 [1.2]    \\ \hline
    \textbf{Solar (GW)}                &  1.8       & 4.7        & 8.3        & 12.7       & 18.8             \\  \hline
    \textbf{BESS (GW)}                 &  0.25      & 0.85 [0.7] & 4.4 [4.35] & 8.9 [8]    & 9.6 [9.55]    \\  \hline 
    \textbf{BESS VI (GW)}              &  0         & 0 (0.15)   & 0 (0.3)    & 0 (0.3)    & 0 (0.35)        \\  
                                       &            & [0.15]     & [0.3]      & [0.85]     & [0.05]        \\  \hline
    \end{tabular}
    \caption{Investment results under BC planning (allowing for changes in operation and additional investments when imposing inertia constraints ex-post) [and IC planning].}
    \label{tab:Results-InvestmentInertia}
\end{table}

In Table \ref{tab:Results-InvestmentInertia}, we observe that indeed by investing in only 3 to 7 BESS units of 50 MW each ex-post, we provide sufficient inertia to make the system RoCoF feasible across all policy cases. While these results are of course system- and data-dependent, they show that by installing a relatively small MW amount of virtual inertia providing units, the system can be made feasible, and at a relatively low additional cost. At most this additional cost amounted to 34.9 M\euro, or 1.4 \% of total system cost.

Since this additional ex-post investment cost is relatively low, we can also expect that the unconstrained IC model would not yield a very different capacity mix from the one observed in Table \ref{tab:Results-InvestmentInertia}. And indeed, in terms of total capacity technology - without distinguishing between units that can and cannot provide virtual inertia - that is true. However, the difference occurs in the location of the resources, and in the type (traditional versus VI providing) of resource. Let us focus on the carbon-neutral (100\% clean production) case. Both models, the BC and the IC build 3.5 GW of wind capacity; however, the BC builds this capacity entirely of traditional wind, whereas the IC model prefers to build 2.3 GW of traditional and 1.2 GW of virtual inertia providing wind resources. Since the wind capacity in the BC is considered sunk, it would not make sense to build even more VI wind capacity ex-post to address the inertia need, and therefore 7 BESS VI units are built instead. This means that, especially in the carbon-neutral case, sub-optimal BC planning might lead to a considerable distortion of the optimal capacity mix. 

Moreover, the location of resources also changes. For that, let us focus on the 90\% case and observe BESS location, which is presented in Table \ref{tab:Results-InvestmentInertiaBESS}. First of all, we note that the IC model builds 17 BESS VI units, whereas the BC model only builds 6 ex-post, so the IC model is provides a more robust system in terms of providing inertia\footnote{This observation is true in general for the observed case studies. In all cases observed, the IC model builds either more or the same amount of VI units than the actual BC case.}. This point is further demonstrated by observing the total average system inertia, which is at its minimum of 7.5 seconds in the adjusted BC case, and at 12.1 seconds under IC planning. When co-optimizing GEP and inertia requirements, the inertia-providing resources not only provide the bare minimum of necessary inertia, they are actually used more for other services such as energy arbitrage, thereby creating a win-win situation: the resources are utilized more and for different services; and, more inertia is provided to the system. 
The GW amount that is placed at each node also varies between IC and BC. Moreover, under the IC case, total BESS capacity is 8.85 GW (summing BESS and BESS VI resources), which is lower than the total BESS capacity of 9.2 GW in the BC case (after being made inertia-feasible). Finally, the total cost of the IC case is 1946.5 M\euro, so 12.6 M\euro { }less than the adjusted BC cost. In relative terms, however, this cost difference is small. In conclusion, we can state that while cost-wise there is not a big difference between planning using inertia constraints (IC) or planning without inertia (BC), but imposing it ex-post; however, the IC planning is cheaper, uses less overall capacity and is more robust with respect to providing inertia.

\begin{table}[ht]
    \centering
    \begin{tabular}{l|c|c|c|c|c}
                        &Bus 1       & Bus 4       & Bus 5     & Bus 6     & Average     \\    
                        & (GW)       & (GW)       &(GW)     & (GW)      & Inertia (s)     \\   \hline 
    \textbf{Actual BC}  & 0          & 0.05        & 0.35      & 8.5       & 7.5     \\  
                        &            & + 0.15VI    & + 0.15 VI &           &    \\  \hline
    \textbf{IC}         & 0          & 0           & 0.15      & 7.95      & 12.1 \\  
                        & + 0.05 VI  & + 0.4 VI    & + 0.2 VI  & + 0.2 VI  &   \\  \hline
    \end{tabular}
    \caption{BESS and BESS VI investment results in GW by bus and total average system inertia in seconds for the 90\% clean production case.}
    \label{tab:Results-InvestmentInertiaBESS}
\end{table}


The main take-aways of section \ref{subsec:ResultsInertia} are: without the introduction of technologies that can provide virtual inertia, the power system cannot exceed a certain threshold of RES production and provide the necessary system inertia at the same time; therefore, a substantial amount of research, testing and prototyping of virtual inertia providing technologies - and doing it quickly - is necessary to achieve ambitious carbon-neutral policy goals; the fraction of VI providing capacity necessary in a power system is relatively low, and hence, disregarding inertia constraints in planning models might not greatly distort the optimal capacity mix - it might, however, within the same type of technology impact the distribution between traditional and VI providing units; finally, even under simplified (BC) planning, additional VI investments are crucial to guarantee system stability and finally, considering inertia in GEP (IC) can lead to a cheaper and more robust system.

\subsection{Sensitivity regarding Reactive Power Constraints}
\label{subsec:ResultsReactive}

In this section we quantify the importance of taking into account the full AC-OPF (via SOCP) and reactive power constraints in generation expansion planning. To that purpose, we compare the base case (BC) model, which uses a DC-approximation of power flow, to the reactive case (RC) model which uses a SOCP representation.
We furthermore compare these two models twice: without clean production constraints (i.e. $\kappa=0$), so business as usual; and, for a 100\% RES power system (i.e. $\kappa=1$).

\subsubsection{Business as usual: no clean production enforced}
\label{subsubsec:ResultsReactiveBAU}
In Table \ref{tab:Results-Reactive0Clean}, we present the main investment results for the business as usual case for the BC and the RC model where no clean production constraints are enforced.

\begin{table}[ht]
    \centering
    \begin{tabular}{l|c|c}
                                  & BC           & RC \\  \hline 
    \textbf{Total Cost (M\euro)}  & 1260.9       &  1279.7 \\  
    \textbf{Actual Cost (M\euro)} & (1282.5)     &        \\  \hline
    \textbf{CCGT (GW)}            & 1.7          &  2.2  \\  \hline
    \textbf{OCGT (GW)}            & 0.8          &  0.4 \\  \hline
    \textbf{Wind (GW)}            & 2.5          &  2.1  \\  \hline
    \textbf{Solar (GW)}           & 1.8          &  1.6 \\  \hline
    \textbf{BESS (GW)}            & 0.25         &  0.25 \\  \hline
    \textbf{FACTS (units)}        & - (6)        &  4
          \\  \hline
    \end{tabular}
    \caption{Total system cost and investment results of the corrected BC (that allows for additional investments in FACTS and changes in operating decisions) and RC models in a system without clean production constraints.}
    \label{tab:Results-Reactive0Clean}
\end{table}

Before we discuss the results in detail, let us point out one important issue. The power system (and its operation) arising under the DC-OPF constraints in the BC model, does not necessarily have to be feasible under an AC setting. As a matter of fact, if we fix the BC capacity mix and impose the AC-OPF and reactive power constraints ex-post without allowing to build additional FACTS devices, we can observe several things: the initially obtained renewable penetration can no longer be achieved, as thermal generators are dispatched more in order to provide the necessary reactive power. Depending on the renewable penetration, the system can even become infeasible all together.  

In order to fix this problem should it arise, we have to change investment decisions. We allow investing in additional units ex-post BC planning. Computationally speaking: we run the BC model; then, we fix the lower bound on investment variables (assuming that investments that have already been decided are sunk), but we allow for additional investments in generating units and FACTS devices; then, we run the RC model (but with the lower bound on investments as determined by the BC in place). The difference in the objective function between the BC and the constrained RC gives us an idea about the additional cost of making sub-optimal DC generation expansion planning AC feasible. 
Therefore, in Table \ref{tab:Results-Reactive0Clean} we have added, in parenthesis, the actual\footnote{The actual cost would be the objective function value obtained with the RC model when fixing the lower bound of investments to the one obtained by the BC model.} cost of sub-optimal BC planning, and the additional investments in infrastructures that are necessary to make it AC-feasible. So when comparing BC versus RC system costs, the correct number is the actual cost and investments in parenthesis. 

In Table \ref{tab:Results-ReactivePowerGeneration0Clean} we show what technologies are providing reactive power in the 0\% RES system and how much through the whole year. We observe that the gas units account for 43\% of annual reactive power provision, the hydro plant at bus 3 accounted for 13.8\% and the remaining reactive power is provided by the FACTS devices.

\begin{table}[ht]
    \centering
    \begin{tabular}{l|c}
                                  & RC \\  \hline 
    \textbf{CCGT (GVarh)}         &  5019.6  \\  \hline
    \textbf{OCGT (GVarh)}         &  80.3 \\  \hline
    \textbf{Hydro (GVarh)}        &  1618.6  \\  \hline
    \textbf{FACTS (GVarh)}        &  5052.8
          \\  \hline
    \end{tabular}
    \caption{Total annual reactive power generation by technology under RC planning in a system without clean production constraints.}
    \label{tab:Results-ReactivePowerGeneration0Clean}
\end{table}

The cost of sub-optimal planning under the BC is only 2.8 M\euro, which is 0.2\% of total system cost. The overall capacity mix looks similar as well with a maximum of 20\% distortion in wind capacity. Solar capacity differs by only 200 MW and the location of this capacity is also slightly different. In the RC approach 1.3 GW are installed at bus 6, the highest demand node, and 0.3 GW are installed at the remote location at bus 8. Whereas under the BC approach all 1.8 GW of solar are installed at bus 6. 
Hence, explicitly modeling reactive power has led to distributing solar power at different locations within the network. 

\subsubsection{Enforcing 100\% RES production}
\label{subsubsec:ResultsReactive100Clean}

We repeat the same study but this time planning for a carbon-neutral system, with a 100 percent ($\kappa$=1) clean production constraint in place. The corresponding investment results can be seen in Table \ref{tab:Results-Reactive100Clean}.

\begin{table}[ht]
    \centering
    \begin{tabular}{l|c|c}
                                  & BC            & RC \\  \hline 
    \textbf{Total Cost (M\euro)}  & 2525.5        & 2589.7 \\  
    \textbf{Actual Cost (M\euro)} & (2604.0)      &        \\  \hline
    \textbf{Wind (GW)}            & 3.5           & 3.4  \\  \hline
    \textbf{Solar (GW)}           & 18.8          & 19.3  \\  \hline
    \textbf{BESS (GW)}            & 9.6           & 9.9  \\  \hline
    \textbf{FACTS (units)}        & - (9)         & 9  
          \\  \hline
    \end{tabular}
    \caption{Total system cost and investment results of the corrected BC (that allows for additional investments in FACTS and changes in operating decisions) and RC models in a 100\% RES power system.}
    \label{tab:Results-Reactive100Clean}
\end{table}

First of all, we note that under the original BC capacity mix, only the hydro storage plant can provide reactive power, which is not enough to satisfy system demand, thereby rendering the capacity mix infeasible in an AC framework. We therefore allow for additional investments ex-post to remedy this problem.

We observe that the original BC planning slightly under-invests in BESS and solar, while building 100 MW additional wind. Since the BC model solves a DC-OPF, where reactive power is not accounted for, no FACTS devices are build. It is important to note that the 2525.5 M\euro { }system cost is for a technology mix that is AC-infeasible. However, when allowing for additional investments in FACTS devices - at each bus - the power system can be made AC-feasible by placing one FACTS unit at each bus yielding a total system cost of 2604.0 M\euro. This means that planning directly with RC could have saved 14.3 M \euro { }(or 0.6 \% of total system cost). The overall system cost and total capacity investments per technology is similar in the RC and adjusted BC cases\footnote{That have been made AC-feasible ex-post.}, however, they are not the same. For example, solar is 0.5 GW lower in the BC case than in the RC case.
When it comes to optimal capacity location throughout the network, the distortion can be significant as shown in Table \ref{Tab:Results-Reactive100CleanNodalInvestments}. Under the DC-OPF hypothesis in BC, we build 9.2 GW of BESS capacity at bus 6, as opposed to the optimal 8.5 GW - a nodal capacity distortion of 8.2\% for BESS. At bus 4, the BESS capacity is about half of what it should optimally be. Bus 4 and 6 are high demand buses, and if congestion occurs in the corresponding transmission line, lacking storage capacity at bus 4 might be a problem for consumers at this bus. 

\begin{table}[ht]
    \centering
    \begin{tabular}{l|c|c|c|c|c|c}
                       & Bus 1         &Bus 4           & Bus 5          & Bus 6          & Bus 8          & Total  \\  
                       & (GW)          & (GW)           & (GW)           & (GW)           & (GW)           & (GW)     \\  \hline
    \textbf{BESS BC}   & 0             & 0.25           & 0.15           & 9.2            &                & 9.6 \\  \hline
    \textbf{BESS RC}   & 0.8           & 0.55           & 0.05           & 8.5            &                & 9.9 \\  \hline
    \textbf{Solar BC}  &               &                &                & 14.9           & 3.9            & 18.8 \\   \hline
    \textbf{Solar RC}  &               &                &                & 17.6           & 1.7            & 19.3 \\  \hline
    \end{tabular}
    \caption{Investments in GW in BESS and Solar technologies per system bus comparing BC and RC planning.}
    \label{Tab:Results-Reactive100CleanNodalInvestments}
\end{table}

In the 100\% RES system reactive power can no longer be provided by traditional thermal generators. The only technologies that can generate reactive power are hydro and FACTs devices. Since we do not allow for an expansion in hydro reservoirs in our model\footnote{This is a realistic assumptions as hydro reservoir resources are limited.}, the FACTS devices account for the biggest share of reactive power generation. Hydro provides 10 \% and FACTS provide 90\% of total annual reactive power in the 100\% RES system. 

\begin{table}[ht]
    \centering
    \begin{tabular}{l|c}
                                  & RC \\  \hline 
    \textbf{Hydro (GVarh)}        &  1459.8  \\  \hline
    \textbf{FACTS (GVarh)}        &  13022.0
          \\  \hline
    \end{tabular}
    \caption{Total annual reactive power generation by technology under RC planning in the 100\% RES system.}
    \label{tab:Results-ReactivePowerGeneration100Clean}
\end{table}

As a take-away, the main conclusions assessing the impact of reactive power constraints are summarized as: the sub-optimal BC planning, due to the DC-OPF representation, can be fixed quite cheaply by installing additional FACTS units, and operating existing units differently, leading to an additional system cost of less than 1 percent of total system cost; and, while the distortion of the total capacity per technology is not large, the impact on the allocation of the capacity can be significant.

\subsection{Full LEGO Planning Results}
\label{subsec:ResultsHolistic}


In this section we only focus on the 100\% RES case for the sake of brevity. Focusing on total system costs, we observe some type of economies of scale when considering inertia and reactive constraints simultaneously in GEP. For example, the LEGO model with a total cost of 2600.6 M\euro { }is 32.2 M\euro { }(or 1.2\%) cheaper than the adjusted BC model which yields a total cost of 2632.8 M\euro. The full LEGO capacity mix fixes inertia and reactive power issues at a cost that is lower than what it would have cost to only fix reactive power issues in sub-optimal BC planning. 

\begin{table}[ht]
    \centering
    \begin{tabular}{l|c|c}
                                  & BC                & LEGO \\  \hline
    \textbf{Total Cost (M\euro)}  & 2525.5            & 2600.6 \\  
    \textbf{Actual Cost (M\euro)} & (2632.8)          &        \\  \hline
    \textbf{Wind (GW)}            & 3.5               & 1.6  \\  \hline
    \textbf{Wind VI (GW)}         & 0                 & 1.8  \\  \hline
    \textbf{Solar (GW)}           & 18.8 (19.4)       & 19.2  \\  \hline
    \textbf{BESS (GW)}            & 9.6               & 9.85  \\  \hline
    \textbf{BESS VI (GW)}         & 0 (0.55)          & 0  \\  \hline
    \textbf{FACTS (units)}        & - (9)             & 9  
          \\  \hline
    \end{tabular}
    \caption{Total system cost and investment results of the corrected BC (that allows for additional investments and changes in operating decisions) and full LEGO models in a 100\% RES system.}
    \label{tab:Results-Holistic100Clean}
\end{table}

Table \ref{tab:Results-Holistic100CleanInertiaReactive} contains a summary of system inertia and reactive power results for a 100\% RES system. Due to the difference in investments in VI technologies, inertia is provided by hydro and BESS VI in the adjusted BC case, and by hydro and Wind VI under LEGO planning. We observe that LEGO planning provides a system that is more robust with respect to inertia provision, being that total average system inertia is 9.2, whereas the adjusted BC yields only 7.5 - the RoCoF minimum - system inertia.
With respect to reactive power generation, in the 100\% RES system only hydro and FACTS can provide reactive power. While the total reactive power generation is provided as 10\% hydro and 90\% FACTS in the adjusted BC case, under LEGO planning this shifts to 6.7\% hydro and 93.3\% FACTS. 

\begin{table}[ht]
    \centering
    \begin{tabular}{l|c|c}
                                  & Adjusted BC            & LEGO \\  \hline
    \textbf{Average Inertia (s)}  & 7.5 (Hydro \& BESS VI) & 9.2 (Hydro \& Wind VI) \\   \hline
    \textbf{Hydro (GVarh)}        & 1456.3            & 957.0  \\  
    \textbf{FACTS (GVarh)}        & 12946.8           & 13253.8  
          \\  \hline
    \end{tabular}
    \caption{Average system inertia in seconds, and reactive power generation in GVarh of the adjusted BC (where ex-post changes in investment and operation decisions are allowed) and LEGO models in a 100\% RES system.}
    \label{tab:Results-Holistic100CleanInertiaReactive}
\end{table}

As observed in previous case studies, the total capacity mix (ex-post) is not very different from the LEGO mix. There is only a 600 MW difference (over 33 GW of total capacity) between the two mixes.
However, the devil lies in the details. First of all, inertia is provided from 1.8 GW of Wind VI capacity in the LEGO model, whereas it is provided by only 0.55 GW of BESS VI in BC. That is a drastic difference in terms of which technology provides inertia in a power system. 
Resources are also located differently, for example the LEGO model places 0.55 GW of BESS at bus 1, whereas BC does not place capacity at bus 1 at all. Moreover, even though BC has a higher total installed capacity, the mix is less efficient in terms of total cost. Finally, the LEGO model yields a power system that is more robust with respect to the provision of system inertia.

\section{Conclusions}
\label{sec:Conclusions}

In this paper we have proposed a novel low-carbon expansion generation optimization (LEGO) model that simultaneously accounts for unit commitment constraints, an SOCP approximation of the AC optimal power flow, and introduces inertia requirements via RoCoF for both synchronous generators and virtual inertia providing units. From our case studies, we conclude the following: first, without explicitly accounting for inertia requirements in generation expansion planning, the obtained capacity mix is incapable of satisfying both RoCoF and high renewable targets at the same time. As a consequence, there is a specific threshold of renewable penetration, i.e., 80\% in our case study, that cannot be exceeded unless inertia requirements are explicitly accounted for in GEP. 

If additional investments in generating units, and in particular, in virtual-inertia providing units are permitted ex-post, the power system can be made RoCoF-feasible at a relatively low cost. The difference in total system cost between a basic planning approach that is made RoCoF-feasible ex-post by allowing additional investments, and the costs of the optimal capacity mix that accounted for inertia from the start is less than 1\%. However, while the impact in total costs might be modest, there can occur significant distortions regarding the location of the resources, and the sub-optimal mix is prone to invest in fewer inertia-providing units and is therefore less robust than the optimal mix.

Reactive power constraints might not be an issue in current bulk power systems; however, in low-inertia grids with high renewable penetration disregarding reactive power constraints in GEP quickly renders an AC-infeasible capacity mix. Allowing for additional investments in FACTS remedies infeasibility problems at a low cost. However, again we observe that disregarding reactive power constraints in GEP leads to a distortion in the allocation of resources, which in turn might have a wider impact on optimal transmission expansion planning. Assessing the impact of generation and transmission co-planning while accounting for inertia and AC-OPF constraints is a topic for future research.

\section*{Acknowledgements}
\label{sec:Acknowledgements}

The authors would like to thank Uros Markovic for helpful comments and suggestions. S. Wogrin also wants to acknowledge MIT LIDS for hosting her research visit and the Jos\'{e} Castillejo grant awarded by the Spanish Ministerio de ciencia, innovaci\'{o}n y universidades.

\newpage
\allowdisplaybreaks
\section*{Nomenclature}

Indices:

\begin{longtable}{ l l }
 $p$                    & Time periods (usually hours)  \\ 
 $rp$                   & Representative periods (usually days)  \\ 
 $k$                    & Time periods within a representative period  \\ 
 $\Gamma(p,rp,k)$       & Mapping of periods with representative periods $rp$ and $k$  \\  
 $g$                    & Generating units \\  
 $t(g)$                 & Subset of thermal generation units \\
 $s(g)$                 & Subset of storage generation units \\
 $r(g)$                 & Subset of renewable generation units \\ 
 $v(g)$                 & Subset of units that provide virtual inertia \\
 $facts(g)$             & Subset of FACTS as reactive power source \\
 $i,j,ii$               & Bus of transmission network \\
 $iws$                  & Transmission busses without slack bus \\
 $c$                    & Circuit in transmission network \\
 $ijc(i,j,c)$           & Transmission line connecting nodes $i$,$j$ with $c$ \\
 $line(i,j)$            & Indicates if a line exists between nodes $i$ and $j$   \\
 $gi(g,i)$              & Generator $g$ connected to node $i$ \\
 
\end{longtable}

Parameters:

\begin{longtable}{ l l }
 $D^P_{rp,k,i}$         & Active power demand (GW)  \\ 
 $D^Q_{rp,k,i}$         & Reactive power demand (GW)  \\ 
 $\eta^{DIS}_{g}$       & Discharge efficiency of unit (p.u.)  \\
 $\eta^{CH}_{g}$        & Charge efficiency of unit (p.u.)  \\
 $B_i$                  & Susceptance connected at bus $i$ (p.u.)  \\
 $B_{i,j,c}$            & Line susceptance (p.u.) \\
 $Bc_{i,j,c}$           & Branch charging susceptance (p.u.) \\
 $G_i$                  & Conductance connected at bus $i$ (p.u.)  \\
 $G_{i,j,c}$            & Line conductance (p.u.) \\
 $SB$                   & Base power (MVA)  \\
 $R_i$                  & Tan(arccos(pf)) = Q/P at bus $i$ (p.u.) \\
 $W^{RP}_{rp}$          & Weight of the representative period (h) \\
 $W^{K}_{k}$            & Weight of each $k$ within the representative period (h) \\
 $C^{ENS}$              & Cost of energy non-served (M\euro/GWh) \\
 $C^{SU}_g$             & Start-up cost of unit (M\euro) \\
 $C^{UP}_g$             & Commitment cost of unit (M\euro/h) \\
 $C^{VAR}_g$            & Variable cost of energy (M\euro/GWh) \\
 $C^{OM}_g$             & Operation and maintenance cost (M\euro/GWh) \\
 $C^{INV}_g$            & Investment cost (M\euro/GW/y) \\
 $C^{RES+}$             & Reserve-up cost (p.u.) \\
 $C^{RES-}$             & Reserve-down cost (p.u.) \\
 $RES^+$                & System reserve-up requirement (p.u.)  \\ 
 $RES^-$                & System reserve-down requirement (p.u.)  \\
 $\underline{P}_g$      & Technical minimum of unit (GW) \\
 $\overline{P}_g$       & Technical maximum of unit (GW) \\
 $EU_g$                 & Indicator of existing unit (integer) \\
 $RU_g$                 & Ramp-up limit of unit (GW) \\
 $RD_g$                 & Ramp-down limit of unit (GW) \\
 $MOW$                  & Moving window for long-term storage (h) \\
 $PF_{rp,k,i,r}$        & Renewable profile per unit and node (p.u.) \\
 $\underline{R}_s$      & Minimum reserve of storage unit (p.u.) \\
 $M^{ch/d}_{rp,k,s}$    & Upper bound on charge and discharge (GW) \\
 $InRes_{s,p}$          & Initial reserve (GWh) \\
 $IF_{rp,k,s}$          & Inflows (GWh) \\ 
 $\kappa$               & Minimum clean (s+r) production (p.u.) \\
 $ISF_{i,j,c,ii}$       & Injection Shift Factors (p.u.) \\ 
 $\overline{T}_{i,j,c}$ & Transmission line limit (GW) \\
 $\overline{A}_{i,j,c}$ & Apparent power transfer limit (MVA) \\
 $\Delta$               & Maximum angle difference (rad) \\
 $\overline{X}_g$       & Maximum amount of units to be built (p.u.) \\

\end{longtable}

Variables:

\begin{longtable}{ l l }
 $p_{rp,k,g}$           & Real power generation of the unit (GW)  \\ 
 $\hat{p}_{rp,k,g}$     & Real power generation above the technical minimum (GW) \\
 $q_{rp,k,g}$           & Reactive power generation of the unit (Gvar)  \\ 
 $cs_{rp,k,g}$          & Consumption of the unit (GW)  \\ 
 $pns_{rp,k,i}$         & Power non-served (GW) \\ 
 $f^P_{rp,k,i,j,c}$     & Real power flow of line $ijc$ (GW) \\
 $f^Q_{rp,k,i,j,c}$     & Reactive power flow of line $ijc$ (Gvar) \\
 $so^{cii}_{rp,k,i}$    & Auxiliary $cii$ variable for SOCP formulation (p.u.) \\
 $y_{rp,k,g}$           & Startup decision of the unit (integer)  \\ 
 $z_{rp,k,g}$           & Shutdown decision of the unit (integer)  \\ 
 $u_{rp,k,g}$           & Dispatch commitment of the unit (integer)  \\ 
 $x_{g}$                & Investment in generation capacity (integer)  \\
 $b^{ch/d}_{rp,k,s}$    & Indicator if storage is charging or discharging (binary)  \\  
 $sp_{rp,k,s}$          & Spillages or curtailment (GWh) \\
 $res^+_{rp,k,g}$       & Secondary reserve up allocation (GW)  \\ 
 $res^-_{rp,k,g}$       & Secondary reserve down allocation (GW)  \\ 
 $inter_{p,s}$          & Inter period storage reserve or state of charge (GWh)  \\
 
\end{longtable}

\bibliography{main}

\end{document}